\documentclass[sigconf,nonacm]{acmart}
\usepackage{systeme}
\usepackage{bbm}
\usepackage{tikz}
\usepackage{wrapfig}
\usepackage{caption}
\usetikzlibrary{bayesnet}
\usepackage{dsfont}
\usepackage{comment}
\usepackage{type1cm} %
\usepackage{graphicx} %
\usepackage{xspace} %
\usepackage{balance} %
\usepackage{booktabs} %
\usepackage{multirow} %
\usepackage{amsmath} %
\usepackage[font={bf}, tableposition=top]{caption} %
\usepackage{booktabs}
\usepackage{bold-extra} %
\usepackage[vlined,linesnumbered,ruled,noend]{algorithm2e} %
\usepackage{microtype} %
\usepackage{siunitx} %
\usepackage{xfrac} %
\usepackage{mathtools} %
\usepackage{xspace} %
\usepackage{enumitem}
\PassOptionsToPackage{hyphens}{url} %
\PassOptionsToPackage{bookmarks, pdftex, colorlinks=true, pagebackref=true, backref=page}{hyperref} %
\usepackage{cleveref} %
\PassOptionsToPackage{square,numbers}{natbib} %
\usepackage[hyperpageref]{backref} %
\usepackage{hyphenat} %
\usepackage[show]{chato-notes}

\usepackage[lofdepth,lotdepth]{subfig}
\tikzstyle{detobs} = [det, fill = gray!30]

\newcommand{\spara}[1]{\smallskip\noindent\textbf{#1}}

\newenvironment{squishlist}
{\begin{list}{$\bullet$}
 {\setlength{\itemsep}{0pt}
     \setlength{\parsep}{3pt}
     \setlength{\topsep}{3pt}
     \setlength{\partopsep}{0pt}
     \setlength{\leftmargin}{1.5em}
     \setlength{\labelwidth}{1em}
     \setlength{\labelsep}{0.5em} } }
{\end{list}}

\renewcommand*\backref[1]{\ifx#1\relax \else (Cited on #1) \fi}

\graphicspath{{./images/}}

\newcommand{\subreddit}[1]{\texttt{r/{#1}\xspace}}

\newcommand{\totusers}{1.5M\xspace}

\newcommand{\totactivism}{4.1k\xspace}

\newcommand{\nlocations}{311\xspace}
\newcommand{\propgeoref}{25.7\%\xspace}

\newcommand{\sym}{\ensuremath{S}\xspace}

\newcommand{\activated}{activated\xspace}
\newcommand{\activation}{activation\xspace}
\newcommand{\awareness}{awareness\xspace}

\newcommand{\K}{\ensuremath{826}}
\newcommand{\vsym}{\ensuremath{\text{\textit{var}}(\sym)}\xspace}

\usepackage{placeins}

\AtBeginDocument{%
  }

\begin{document}

\title{Causal Modeling of Climate Activism on Reddit}

\author{Jacopo Lenti}
\orcid{0000-0003-2886-7338}
\affiliation{%
  \institution{Sapienza University}
  \city{Rome}
  \country{Italy}
}
\affiliation{%
  \institution{CENTAI}
 \city{Turin}
 \country{Italy}
}
\email{jcp.lenti@gmail.com}

\author{Luca Maria Aiello}
\affiliation{%
  \institution{IT University of Copenhagen}
  \city{Copenhagen}
  \country{Denmark}}
\email{luai@itu.dk}

\author{Corrado Monti}
\affiliation{%
  \institution{CENTAI}
  \city{Turin}
  \country{Italy}
}
\email{corrado.monti@centai.eu}

\author{Gianmarco \\ De Francisci Morales}
\orcid{0000-0002-2415-494X}
\affiliation{%
 \institution{CENTAI}
 \city{Turin}
 \country{Italy}}
\email{gdfm@acm.org}

\renewcommand{\shortauthors}{Lenti et al.}

\begin{abstract}
Climate activism is crucial in stimulating collective societal and behavioral change towards sustainable practices through political pressure.
Although multiple factors contribute to the participation in activism, their complex relationships and the scarcity of data on their interactions have restricted most prior research to studying them in isolation, thus preventing the development of a quantitative, causal understanding of why people approach activism.
In this work, we develop a comprehensive causal model of how and why Reddit users engage with activist communities driving mass climate protests (mainly the 2019 Earth Strike, Fridays for Future, and Extinction Rebellion).
Our framework, based on Stochastic Variational Inference applied to Bayesian Networks, learns the causal pathways over multiple time periods.
Distinct from previous studies, our approach uses large-scale and fine-grained longitudinal data (2016 to 2022) to jointly model the roles of sociodemographic makeup, experience of extreme weather events, exposure to climate-related news, and social influence through online interactions.
We find that among users interested in climate change, participation in online activist communities is indeed influenced by direct interactions with activists and largely by recent exposure to media coverage of climate protests.
Among people aware of climate change, left-leaning people from lower socioeconomic backgrounds are particularly represented in online activist groups.
Our findings offer empirical validation for theories of media influence and critical mass, and lay the foundations to inform interventions and future studies to foster public participation in collective action.

\end{abstract}

\begin{CCSXML}
<ccs2012>
   <concept>
       <concept_id>10002951.10003260.10003282.10003292</concept_id>
       <concept_desc>Information systems~Social networks</concept_desc>
       <concept_significance>500</concept_significance>
       </concept>
   <concept>
       <concept_id>10003120.10003130.10003131.10011761</concept_id>
       <concept_desc>Human-centered computing~Social media</concept_desc>
       <concept_significance>500</concept_significance>
       </concept>
   <concept>
       <concept_id>10002950.10003648.10003662</concept_id>
       <concept_desc>Mathematics of computing~Probabilistic inference problems</concept_desc>
       <concept_significance>300</concept_significance>
       </concept>
 </ccs2012>
\end{CCSXML}

\ccsdesc[500]{Information systems~Social networks}
\ccsdesc[500]{Human-centered computing~Social media}
\ccsdesc[300]{Mathematics of computing~Probabilistic inference problems}

\keywords{Social Media, Collective Action, Opinion Dynamics, Climate Change}

\maketitle

\tikzstyle{obs} = [circle,fill=white,draw=black,inner sep=1pt,
    minimum size=25pt, font=\fontsize{8}{10}\selectfont, node distance=0.5,fill=gray!25 ]
\tikzstyle{latent} = [circle,fill=white,draw=black,inner sep=1pt,
    minimum size=25pt, font=\fontsize{8}{10}\selectfont, node distance=0.5]
\tikzstyle{const} = [circle,fill=white,draw=white,inner sep=1pt,
    minimum size=25pt, font=\fontsize{8}{10}\selectfont, node distance=0.5]

\begin{figure*}[t]
\vspace{-\baselineskip}
\begin{center}
      \begin{minipage}{0.65\textwidth}
        \hspace{-0.8cm}
      \resizebox{0.9\linewidth}{!}{
      \begin{tikzpicture}
        \node[latent]	(SDu)     {$D$};
        \node[obs, right=1.6cm of SDu, yshift = 1.1cm]     (SUBlt)    {$P_{L}$};
        \node[obs, left=1.0cm of SUBlt, yshift = 0.9cm]     (POP)    {$N_{sub}$};
        \node[obs, left=0.4cm of SUBlt, yshift=2cm]        (SDsub)   {$D_{sub}$};
        \node[obs, right=5cm of SUBlt]        (SUBst)     {$P_{S}$};
        \node[latent, right=4.8cm of SDu]     (O)     {\sym};
        \node[obs, right=1.2cm of SDu, yshift=-1cm]     (ACTlt)     {$E_{L}$};
        \node[obs, right=4cm of ACTlt]     (ACTst)     {$E_{S}$};
        \node[obs, left=0.2cm of O, yshift=-2.2cm]     (Mlt)     {$M_{L}$};
        \node[obs, right=5.9cm of O]     (A)     {$A$};
        \node[obs, right=0.8cm of SUBst]     (INst)     {$I$};
        \node[obs, left=2cm of A, yshift=-2.2cm]     (Mst)     {$M_{S}$};
        
        \edge {SDu} {SUBlt}
        \edge {SDu} {O}
	\path (POP) edge [->, >={triangle 45}, out=-10, in=160] (SUBlt) ;%
	\path (POP) edge [->, >={triangle 45}, out=0, in=160] (SUBst) ;%
	\path (SDsub) edge [->, >={triangle 45}, out=-20, in=140] (SUBlt) ;%
	\path (SDsub) edge [->, >={triangle 45}, out=-10, in=140] (SUBst) ;%
	\path (SDsub) edge [->, >={triangle 45}, out=0, in=150] (INst) ;%
        \edge {SUBlt} {SUBst}
        \edge {ACTlt} {SUBlt}
	\path (ACTlt) edge [->, >={triangle 45}, out=10, in=200] (O) ;%
        \edge {ACTlt} {ACTst}
        \edge {O} {SUBst}
        \edge {O} {A}
        \edge {SUBst} {INst}
        \edge {Mlt} {O}
	\path (Mlt) edge [->, >={triangle 45}, out=15, in=220] (A) ;%
	\path (Mst) edge [->, >={triangle 45}, out=10, in=240] (A) ;%
        \edge {ACTst} {SUBst}
	\path (ACTst) edge [->, >={triangle 45}, out=10, in=200] (A) ;%
        \edge {ACTst} {INst}
        \edge {INst} {A}

      \end{tikzpicture}
      }
    \end{minipage}%
    \hspace{-1cm}
    \small
    \begin{minipage}{0.3\textwidth}
        \centering
    \begin{tabular}{@{}ll@{}}
    \toprule
    \textbf{Node Name} & \textbf{Description} \\ \midrule
    $D$       & Demographics of the user \\ 
    $D_{sub}$        & Demographics of the subreddits \\ 
    $N_{sub}$             & Number of users in the subreddits \\ 
    $P_{L}$        & Participation to subreddits (long term) \\ 
    $P_{S}$        & Participation to subreddits (short term) \\
    \sym               & User sympathy for activism at $t_{A-W}$ \\ 
    $E_{L}$        & Reddit Engagement (long term)\\ 
    $E_{S}$        & Reddit Engagement (short term)\\ 
    $M_{L}$          & Media coverage (long term) \\ 
    $M_{S}$          & Media coverage (short term) \\ 
    $A$               & User activated in climate activism groups \\ 
    $I$         & Interactions with activists \\ 
    \bottomrule
    \end{tabular}
\end{minipage}
    \caption{Probabilistic Graphical Model (left) and description of nodes (right). Circles represent stochastic variables; %
    gray nodes are observed and white ones are latent. 
      \label{fig:pgm}}
\vspace{-\baselineskip}
\end{center}
\end{figure*}

\section{Introduction}
The climate crisis is an unprecedented challenge for humanity, as the last decade has been the warmest in over \num{125 000} years~\cite{ipcc2022}.
Climate change is a collective action problem that requires mass coordination due to its global scale and to the interdependency of all agents involved~\cite{ipcc2014}.
In recent years, the younger generations have succeeded in mobilizing millions of people worldwide to draw attention to the climate emergency~\cite{han2020youth}.
These collective events show the centrality of the climate change issue in our societies, and demonstrate that humanity can organize global collective action for a common cause.

Collective action is defined as any action that aims to improve the status, power, or influence of a group~\cite{van2009introduction}.
We call \emph{climate action} the set of collective strategies intended to increase political and social pressure to quickly enact effective climate mitigation measures~\cite{ipcc2022}.
Since climate risks~\cite{emmerling2024multi}, responsibilities~\cite{2022demographics}, and awareness~\cite{doell2024international} are spread unequally in the global population, it is natural that climate activism is also unequally spread.
As such, climate action is the most pristine example of how challenges facing humanity are, first and foremost, a problem of social coordination.
Therefore, investigating the mechanism of collective action and activism is fundamental not only to tackle climate change in itself, but also to understand how to address other global crises---from pandemics to misinformation.

Research in the social sciences has identified different pathways to engagement with collective action, primarily focusing on cost-benefit, collective efficacy, group-based emotions, and social identity theory~\cite{tajfel1978social,tajfel1979integrative}.
So far, the validity of these theories has only been assessed by survey-based studies~\cite{van2013social}, where it is hard to collect heterogeneous and balanced data.
However, the abundance of human behavior data from social media opens a different perspective into these dynamics.
Previous social media studies on climate action have analyzed the changes in climate debate after extreme weather events~\cite{torricelli2023does, roxburgh2019characterising, sisco2017extreme, kirilenko2015people}, political events~\cite{falkenberg2022growing, mavrodieva2019role} or media coverage~\cite{kirilenko2015people}.
However, these studies are either associational or focus on single causal pathways for the phenomena of interest.
Thus, they neglect potentially important interactions effects between the key causal mechanisms involved.
To gain a richer understanding of the processes sparking collective action and compare the effects of different causal mechanisms at play, a \emph{comprehensive causal model} is needed instead.
We contribute to this objective by applying a Bayesian causal approach to analyze the pathways leading people to engage with climate activism on Reddit.

We envision the path leading to collective action as a funnel, where several essential steps filter the overall population toward active mobilization.
We focus on the segment of this selection process that brings users who are aware of the climate change issue to become participants in climate activist communities.
This step is a fundamental filter, as it determines who---among the crowd of aware individuals that are potential targets for mobilization---is motivated enough to take a first perceptible step towards action: participating in an online forum dedicated to climate action.

We identify as our target variable the participation in one of the subreddits affiliated with the organizations that coordinated the September 2019 global climate strikes, mainly \emph{Earth Strike}, \emph{Extinction Rebellion}, \emph{Fridays For Future}, and the \emph{Sunrise Movement} (see \Cref{apx:subreddits}).
The importance of Reddit in this process is demonstrated by the history of the Earth Strike movement: it was founded after a Reddit post, which led to the creation of a subreddit, and then evolved into the organization of the global Earth Stike on 27 September 2019.\footnote{\url{https://www.reddit.com/r/chomsky/comments/d7i2g0/how_a_post_on_rchomsky_helped_start_the_biggest}}

The goal of this work is to \emph{quantify the relative importance of different social and information processes in causing engagement with activist communities on social media.} Based on existing literature, we identify three main potential pathways leading to this type of engagement, which we call \emph{\activation}.
First, the familiarity with extreme weather events~\cite{van2015social, torricelli2023does} and the media coverage around climate change~\cite{kirilenko2015people} affect users' involvement in the climate debate (\textbf{RQ1}).
Second, the sociodemographic features affect the perception of risk due to climate change~\cite{van2015social} (\textbf{RQ2}). 
Lastly, the intention to participate in collective action is affected by social influence~\cite{van2013social,russo2023stranger} (\textbf{RQ3}).
We articulate our three main research questions accordingly:
\begin{squishlist}
\item[\textbf{RQ1}:] Does media coverage about climate and climate action affect \activation in climate activism groups on Reddit, and over which time scale?
\item[\textbf{RQ2}:] Which sociodemographic characteristics affect \activation?
\item[\textbf{RQ3}:] Do the interactions with other activist individuals affect one's \activation?
\end{squishlist}

To answer these questions, we introduce a causal model based on Bayesian Networks, depicted in \Cref{fig:pgm}, that learns multiple key determinants of engagement with activist communities from a ground truth on \activation extracted from large-scale Reddit data.
This is the first data-driven study that considers the interplay between different causal pathways towards collective action in the context of social media.
The pipeline we propose can be flexibly repurposed for the study of other phenomena of collective engagement from online traces.

Our results show that social influence is the main driver of engagement.
Sociodemographic features influence the engagement in climate activism, as they increase the sympathy for the cause and probability of frequenting subreddits where social influence takes place.
Finally, mass media coverage of climate action has a positive effect on the engagement in the short term.

\section{Related Work}
\label{sec:related}

Social media platforms proliferate with discussions around climate change, ranging from promotion of grassroots action to expressions of climate skepticism~\cite{chen2023climate, gaupp2024climate}.
Online communities focused on the climate crisis have become a fertile ground for the development of a generational social identity~\cite{gaupp2024climate}.
Social identity is one of the main pathways for engaging in collective behavioral change~\cite{bamberg2015collective,schulte2020social}, complementing other drivers of action such as rational incentives and purely emotional triggers~\cite{bamberg2015collective}.
These strongly identity-centric communities played a significant role in organizing the global climate protests of 2019, which drawn broad political attention to the climate emergency~\cite{han2020youth}.

The study of community engagement on social media has frequently been approached through classification tasks, emphasizing the power of the different models and features in predicting user engagement~\cite{purohit2011understanding, zhang2017community, massachs2020roots}.
Some studies relied on exact matching or propensity score matching to investigate causal links between interventions and engagement outcomes, such as joining discussion groups~\cite{phadke2021makes, russo2023stranger}.
These studies primarily focus on endogenous processes within social media, and consider treating explanatory variables as uniform sets, without attempting to model the complex causal pathways connecting them.

Conversely, research on public engagement with climate issues has largely concentrated on external climate-related events and their correlation with social media responses~\cite{roxburgh2019characterising, sisco2017extreme, kirilenko2015people, torricelli2023does, falkenberg2022growing, mavrodieva2019role}.
The relationship between sociodemographic factors and climate activism has been explored mostly through questionnaires~\cite{wang2018analysis, islam2013investigation,van2015social} and connected the personal traits of climate activists to existing psychological theories of collective action~\cite{neas2022young, sabherwal2021greta}.

Recent literature on Probabilistic Generative Agent-Based Models (PGABMS) proposed the use of Bayesian Networks to leverage observed longitudinal data for learning the causal mechanisms to be encoded into those models~\cite{lenti2024likelihood}, including for example the backfire effect~\cite{monti2020learning, lenti2024variational}. Similarly, probabilistic generative frameworks have been used to model community engagement to study
dynamics such as social influence, the echo chamber effect, or political ideology~\cite{minici2022cascade, monti2021learning}.

\section{Model}
\label{sec:framework}

We ground our operationalization of involvement with activist communities---the \emph{\activation} outcome we aim to build a causal model for---in the social psychology theory of collective action by \citet{van2013social}.
According to the theory, before participating in collective action, an individual needs to be 1) a sympathizer for the cause, 2) a target of mobilization attempt, 3) motivated to participate, and finally 4) a participant.
We focus on the dynamics leading %
to step 2: how people who are \emph{aware} of climate change become \emph{active participants} (or \emph{activated} for brevity) in online groups of climate activists where upcoming mobilization events are discussed, and can concretely participate to them if properly motivated~\cite{van2013social}.
This operationalization is supported by the view that activist groups are ``opinion-based groups''~\cite{bliuc2007opinion,mcgarty2009collective}, whose shared identity is based on common values and beliefs towards societal issues.
These opinion-based groups are primarily formed to convert broad beliefs into collective action, and indeed, having a social identity aligned with such a group is strongly associated with activism~\cite{schulte2020social}.
Next, we define the features that constitute our model, the causal relationships between them, and how we estimate the model's parameters.

\subsection{Features}
\spara{Activation.}
We consider our initial population of \emph{aware} individuals to be Reddit users who mentioned the words \emph{``climate change''} or \emph{``global warming''} in any message between January 2016 and January 2022, which results in \totusers users.
The use of these locutions indicates awareness of the climate issue by the time of posting.
We define the \emph{observation period} for each of these users to be between one year before and two years after their first use of such locutions.
Note that this group of Reddit users is the whole population for which we wish to estimate the causal effects.

We select the groups of climate activists as the subreddits directly connected to political groups or movements that took part in the organization of global climate strikes.
In particular, we take as reference the ``Global Week For Future'' event, which took place from 20 to 27 September 2019, and saw the participation of about six million people worldwide.\footnote{\url{https://www.theguardian.com/environment/2019/sep/27/climate-crisis-6-million-people-join-latest-wave-of-worldwide-protests}}
We find six groups that fulfill this criterion, which we describe in \Cref{apx:subreddits}. 
The three main ones are \emph{Extinction Rebellion}, \emph{Earth Strike}, and \emph{Fridays For Future}, which account for more than 92\% of the \activated users we observe.
We say that an \emph{\activation} happens when a user in our population posts their first message in one of these six communities.
Standing on Social Identity Theory, we identify this moment as the first observable instance of their identification with a group characterized by a specific collective identity~\cite{bamberg2015collective}.
We consider a user \emph{\activated} if they do post such a message during their observation period (after we observe their \awareness), and we represent this outcome as a binary variable~$A$, equal to $1$ for \activated users and $0$ for the others.
In total, we obtain \totactivism \activated users.
To estimate our causal model, we take an equally-sized sample of non-\activated users.
To ensure basic comparability of the two groups, we verify that the difference between the Reddit engagement, measured as average number of comments per week, is not significant.

\spara{Temporal dynamics.}
To discern the temporal scale of the causal determinants of \activation, let us define $t_A$ as the time we observe an \activation, or the end of the observation period for non-\activated users.
We consider two distinct time periods.
The \emph{short-term} period contains the last week before $t_A$, i.e., the interval $[t_{A-W}, t_A]$.
The \emph{long-term} one, instead, starts one year before $t_A$ (indicated with $t_0$) and ends five weeks before $t_A$, i.e., $[t_0, t_{A-5W}]$.
\Cref{fig:temporal} illustrates this temporal split.
We leave a four-week gap to separate more clearly the effects between these two periods.
However, removing the gap by extending the long-term period to $t_{A-W}$ does not change our results (see \Cref{apx:robustness}).

\usetikzlibrary{decorations.pathreplacing} %

\begin{figure}%
\vspace{-\baselineskip}
\begin{center}
      \resizebox{0.9\linewidth}{!}{
\begin{tikzpicture}
    \draw[thick] (0, 0) -- (8, 0);

    \filldraw[black] (0, 0) circle (2pt);
    \filldraw[black] (5, 0) circle (2pt);
    \filldraw[black] (7, 0) circle (2pt);
    \filldraw[black] (8, 0) circle (2pt);

    \node[below] at (0, 0) {$t_0$};
    \node[below] at (5, 0) {$t_{A-5W}$};
    \node[below] at (7, 0) {$t_{A-W}$};
    \node[below] at (8, 0) {$t_A$};

    \draw [decorate, decoration={brace, amplitude=10pt}] (0, 0.3) -- (5, 0.3) node[midway, above=10pt] {long-term};
    \draw [decorate, decoration={brace, amplitude=10pt}] (7, 0.3) -- (8, 0.3) node[midway, above=10pt] {short-term};
\end{tikzpicture}
}
\end{center}
\caption{Temporal dynamics of the model. The final time
$t_A$ is the \activation time, or the end of the observation period for non-\activated users. The first three points corresponds to one year, five weeks, and one week before $t_A$. \label{fig:temporal}}
\end{figure}
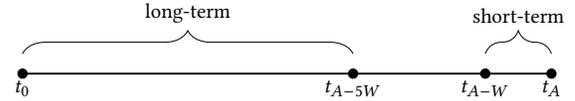

\spara{Sympathy.}
To distinguish how factors in the long- and short-term periods affect activism, we introduce a measure of how likely an individual is to join an activist group at $t_{A-W}$, the beginning of the short-term period.
We call this quantity the \emph{sympathy for the cause} (\sym), referring to the social psychology concept defined by~\citet{van2013social} as the first step towards mobilization.
As sympathy cannot be observed directly, we use a continuous latent variable whose estimation strategy is explained at the end of this section.

We can thus divide the factors we wish to study into long-term ones, whose effect on $A$ is mostly mediated by \sym, and short-term ones, that affect $A$ after $t_{A-W}$.
In particular, we consider:
\begin{itemize}[leftmargin=*]
    \item the user's sociodemographics characteristics $D$, which precede all other factors;
    \item the participation of a user to other subreddits during the long-term ($P_{L}$) or short-term ($P_{S}$) period;
    \item the media coverage of climate-related news, divided into long-term ($M_{L}$) and short-term ($M_{S}$);
    \item interactions with \activated users, that we indicate with $I$;
    \item the overall engagement with the Reddit platform (an essential confounder for our target variable $A$), denoted by $E_{L}$ and $E_{S}$.
\end{itemize}

The rest of this section explains the operationalization and the assumptions behind each of these factors, and how they are linked.
\Cref{fig:pgm} depicts the resulting causal graph.

\spara{Sociodemographic scores.}
We characterize users on four sociodemographic axes based on the methodology developed by~\citet{waller2021quantifying}: affluence (poor to rich), partisanship (left to right), gender (male to female), and age (young to old).
For each axis, they assign a score to each of the 10k most popular subreddits in their data set.
To integrate this methodology into our probabilistic model, we indicate such subreddit scores as $D_{sub}$, consisting of a $K \times 4$ matrix, where $K$ is the number of subreddits we consider.
Then, we represent the sociodemographics of a given user as a 4-dimensional vector $D$ that we estimate as a latent variable, and then we standardize as z-scores.
Thus, the 4 values of $D$ for a user represent their likelihood of belonging to a specific combination of demographic groups: poor-rich, left-right, male-female, and young-old.

\spara{Subreddit participation.}
To estimate the user sociodemographics latent variable $D$, we use user participation in the set of $K$ subreddits that we consider, in the long-term period.
In other words, we assume that the observed subreddit participation is a noisy realization of a user's real sociodemographics.
To determine this set, we first restrict our attention to the \num{1000} most popular subreddits in our dataset.
We intersect it with those considered by \citet{waller2021quantifying}, thus resulting in $K=\nolinebreak\K$ subreddits with sociodemographic scores.
For each user, we define two observed variables 
$P_L$ and $P_S$, representing their participation in these subreddits in the long and short term, respectively.
For a given user, each of those two variables is a binary vector with dimension $K$, with values set to $1$ for the subreddits where the user posted a message during the specified period. 

\emph{Subreddits attributes.}
As a factor affecting the participation of users in subreddits, we consider not only the sociodemographic scores $D_{sub}$, but also the subreddit's overall popularity.
We quantify popularity as the z-score of the log-transformed number of users in our dataset who participate in the subreddit by posting or commenting.
We denote this variable with $N_{sub}$.

\spara{Media coverage.}
We wish to investigate the role of media coverage in the mobilization of activists.
Furthermore, we wish to account for the geographic location of Reddit users and their proximity to the news, when possible.
To do so, we use the Global Database of Events, Language, and Tone (GDELT),\footnote{\url{http://www.gdeltproject.org}}
an open database of geolocated news articles.
We consider three themes that are directly related to climate change: ``climate'', ``climate action'', and ``natural disasters''.
The first theme includes general news about the climate; the second focuses on climate activism (e.g., protests or campaigns); the third involves climate hazards, such as floods or heat waves, that are often a consequence of climate change~\cite{newman2023global}.
Since Reddit is predominantly English-speaking and US-centered,\footnote{More than half of its traffic comes from the US or Canada: \\ \url{https://www.similarweb.com/website/reddit.com}}
in order to consider a homogeneous media landscape, we only consider news located in the United States or Canada.
For each administrative area (i.e., US state or Canadian province), we measure the weekly media attention on a theme as the fraction of news published in that area on that theme.
This way, we obtain a vector of three variables for any period and area.

\emph{User georeferencing.}
\label{sec:georeference}
We wish to associate, whenever possible, Reddit users in our population to their home location.
To do so, we rely on a method developed by~\citet{balsamo2019firsthand}, which leverages user participation in location-based subreddits.
To identify such subreddits, we first gather all the subreddits whose name matches with an existing location and clean them manually to filter out spurious matches~\cite{mejova2021youtubing}, thus obtaining a list of \nlocations location-based subreddits in the US and Canada.
Then, we assign a user to an area if all the location-based subreddits they wrote into within our observation period belong to that area.
This way, we can assign a location in the US and Canada to \propgeoref of our population.
The proportion of georeferenced users is similar between \activated users and our sample of non-\activated users (24.9\% vs. 26.6\%).
Furthermore, the number of users assigned to each area is proportional to the population of that area, thus suggesting strong reliability and homogeneous distribution (\Cref{apx:georef}).
Now, we can associate the location-based media coverage with each user.
For all the users without location, we use the median media coverage per week across all locations, as a standard imputation strategy for missing values. 
Therefore, we obtain for each user two variables, $M_{L}$ and $M_{S}$, respectively for the long-term and the short-term period.
Each variable is a vector of three elements, corresponding to the three news themes (``climate'', ``climate action'', and ``natural disasters'').
$M_{L}$ and $M_{S}$ are standardized to their z-scores.

\spara{Interactions.}
One of our objectives is to study how direct interactions with activists affect the chances of \activation.
A direct interaction $(u,v)$ occurs when a user $u$ leaves a comment below a comment or submission by $v$.
We collect all direct interactions between \activated and non-\activated users in both directions ($(u,v)$ and $(v,u)$).
We define a binary variable $I$ to express whether a given user had at least one interaction with an activist user in the short-term period.
Since we expect capturing this kind of causal relationship over long periods to be fickle, we restrict ourselves to considering only the effect of short-term interactions.

\spara{Reddit engagement.}
Finally, as an essential confounder for the participation in activist subreddits, we consider the overall Reddit engagement of a user.
More active users are in fact necessarily more likely to interact with other users and visit new subreddits.
We measure this engagement as the logarithm of the average number of comments per week in each period for a given user.
Thus we denote $E_L$ their engagement in the long-term period, $E_S$ short-term one.
The variables are standardized to their z-scores.

\subsection{Structure}
Next, we describe the structure of the causal model, explaining each causal relationship between variables.
\Cref{fig:pgm} depicts the Directed Acyclic Graph (DAG) associated with the causal model.
We strive to include in the model all those relations that we cannot exclude through reasoning, in order to
control the confounders and close possible backdoor paths~\citep{pearl2009causality}.
Then, by fitting the model, we can assess which relationships are not substantiated by data.

\spara{Independent variables.}
Data of media coverage, $M_{L}$ and $M_{S}$, comes from external sources, and we consider them as independent of user activity on Reddit.
For this reason, these variables do not have any parents, and are directly observable.
The observed data about subreddits---namely, their demographic scores $D_{sub}$ and their overall popularity $N_{sub}$---are also unaffected by user variables.
Regarding the user, since the sociodemographics are intrinsic and predate all other factors, $D$ does not have any parents.

We consider $E_L$, the observed engagement with Reddit in the long-term period, to be independent of all other factors.
Similarly, we consider its short-term period analogue, $E_S$, to be affected only by its long-term predecessor ($E_L \rightarrow E_S$).
As such, we model it as a Gaussian variable dependent on $E_L$, whose parameters (coefficient, bias, and variance) are to be estimated
\begin{equation}
E_{S} \mid E_{L} \sim \mathcal{N}(\beta_{E1} E_{L} + \beta_{E0}, \theta_E).
\label{eq:1}
\end{equation}

\spara{Long-term participation.}
The demographics of a user $D$ influence the subreddits they participate in, and thus $D \rightarrow P_{L}$.
This effect interacts with the sociodemographic scores of the subreddit $D_{sub}$; 
for instance, left-leaning users are more likely to participate in subreddits categorized as left-leaning.
$P_{L}$ is also influenced by $N_{sub}$ because, in general, users are more likely to join popular subreddits, so we include $N_{sub} \rightarrow P_{L}$.
Moreover, more engaged users are more prone to write in any subreddit, and we model this effect with $E_L \rightarrow P_L$.
We model these causal relations with a logistic function: the participation in a subreddit is a Bernoulli random variable whose probability is given by
\begin{equation}
P(P_{L} \mid D_{sub}, N_{sub}, D, E_{L}) = \sigma(\beta_{P1} D_{sub} \cdot D + \beta_{P2} N_{sub} + \beta_{P3} E_{L} + \beta_{P0}),
\label{eq:2}
\end{equation}
where $\sigma(x) = \frac{1}{1 + e^{-x}}$ and $\beta_{\bullet}$ are coefficients to be estimated.

\spara{Sympathy.}
As mentioned, $S$ denotes the sympathy for the cause---how likely is an individual to join an activist group---at $t_{A-W}$.
As such, it is influenced by the demographics $D$, the Reddit engagement $E_L$, and the media coverage $M_L$, that all precede it temporally.
In fact, given that a person's sociodemographics shape their experience of the world, our model assumes that the influence of sociodemographic $D$ on $A$ is mediated by their internal belief \sym.
Thus, we represent $S$ as a continuous variable, drawn from a normal distribution defined by
\begin{equation}
\sym \mid D, E_{L}, M_L \sim \mathcal{N} (\beta_{S1} D + \beta_{S2} E_{L} + \beta_{S3} M_{L}, \theta_S).
\label{eq:sympathy}
\end{equation}

\spara{Short-term participation.}
Finally, we include the effect of community retention~\cite{zhang2017community} that models the fact that individuals are more likely to write on a subreddit if they contributed to them in the past ($P_{L} \rightarrow P_{S}$).
We hypothesize the same effect of popularity and demographics that we consider in the long-term to be possible in the short term via $N_{sub} \rightarrow P_{S}$ and $D_{sub} \rightarrow P_{S}$.
Analogously, it is reasonable that subreddit participation is affected by the overall Reddit engagement in this period ($E_S \rightarrow P_{S}$).
Finally, since $P_S$ is measured in the short-term period $[t_{A-W}, t_A]$, we hypothesize that it could be affected by the sympathy for the cause $S$, which reflects a measure at $t_{A-W}$.
To avoid estimating the effect of sympathy on each of the $K=\K$ subreddits we consider, which would be prone to overfitting and hard to interpret, we model this relation by aggregating subreddits over the four sociodemographic axes.
Thus, we obtain a term $\beta_{p1} S \cdot D_{sub}$, where $\beta_{p1}$ is a four-element vector that expresses the propensity of a user with high sympathy for the activism cause to participate in subreddits with a given demographic profile
\begin{equation}
\begin{gathered}
P(P_{S} \mid D_{sub}, \sym, P_{L}, N_{sub}, E_S) = \\
\sigma(\beta_{p1} \sym \cdot D_{sub} + \beta_{p2} P_{L} + \beta_{p3} N_{sub} + \beta_{p4} E_S + \beta_{p0}).
\label{eq:P_S}
\end{gathered}
\end{equation}

\spara{Interactions.}
Users who are more engaged on Reddit have a higher probability of interacting (by chance) since interactions are induced by comments.
Hence, we add the relation $E_{S} \rightarrow I$.
The probability of interacting with some \activated user depends on the specific subreddit frequented, which is captured by the causal relationship $P_{S} \rightarrow I$.
In particular, we model this effect via the sociodemographic scores of the subreddits, assuming that the probability of interaction depends on the sociodemographic scores of the subreddits rather than the specific identity of the subreddit.
This design choice enables a more robust model: since the number of subreddits $K$ is quite large, estimating so many parameters would be prone to overfitting.
Therefore, we introduce the effects $D_{sub} \rightarrow I$.
Combining these factors, we obtain
\begin{equation}
P(I \mid D_{sub}, P_{S}, E_{S}) = \sigma(\beta_{I1} P_{S} \cdot D_{sub} + \beta_{I2} E_{S} + \beta_{I0})
\label{eq:5}
\end{equation}

\spara{Activations.}
The literature on opinion dynamics, social influence, and community engagement tells us that interacting with people can have a direct impact on the probability of \activation~\cite{friedkin1990social,friedkin1998structural,russo2023stranger}, and we capture this social influence via $I \rightarrow A$.
Next, we express the role of $S$ by introducing the relation $S \rightarrow A$.
Since $S$ represents how likely a user is to join an activist group at $t_{A-W}$, this relation connects $S$ to the actual activation at $t_A$.
The sociodemographics $D$ precede $S$, temporally and causally, so their effect on $A$ is mediated by $S$, with no direct effect on $A$.
Then, to assess the role of media coverage on activation we introduce $M_S \rightarrow A$, expressing how the media in the week before the activation affects $A$.
The media coverage in the long-term period, $M_L$, can affect the sympathy (at $t_{A-W}$, the end of that period) but could also have a delayed, direct effect on $A$ ($M_L \rightarrow A$)~\cite{torricelli2023does}.
Finally, the effect $E_{S} \rightarrow A$ acts as a confounder, since the \activation corresponds to visiting a new subreddit, and users who are more engaged with Reddit in this period may activate by chance.
\begin{equation}
\begin{gathered}
P(A \mid I, M_{S}, M_{L}, \sym, E_{S}) = \\
\sigma(\beta_{A1} \sym + \beta_{A2} I + \beta_{A3} M_{S} + \beta_{A4} M_{L} + \beta_{A5} E_{S} + \beta_{A0}).
\end{gathered}
\label{eq:activation}
\end{equation}

\subsection{Parameter estimation}
\label{subsec:estimation}
To estimate the model's parameters and latent variables, we fit the model to our Reddit data using Stochastic Variational Inference~\cite{hoffman2013stochastic}.
During the hyperparameter tuning phase, we select: ($i$) the Adam optimizer with a learning rate equal to $0.05$ to maximize the evidence lower bound (ELBO), and ($ii$) a Normal distribution with zero mean and variance $0.1$ as the variational distribution.
To discard results reached in suboptimal local minima of the parameter space, we run the optimization with $10$ random restarts and choose the restart with the best fit to the observed activations.
We measure the fit via accuracy, namely the average fraction of correct predictions of $100$ posterior predictive samples of $A$.
Recall that our goal is to assess causal relations: we are not interested in reaching the highest accuracy, but only in obtaining the most reliable model that respects the previously defined causal assumptions.

While the hyperparameter selection is fully data-driven, the choice of the prior distribution encodes our existing knowledge of the generative process in the context of Reddit.
In particular, we focus on the latent sympathy variable \sym.
The variance \vsym of the prior of \sym regulates the level of stochasticity of the user's sympathy---namely the extent to which sympathy is shaped by individual factors lying outside our model.
At one extreme, \vsym is zero, which makes \sym a deterministic function of its parents.
This assumption is unrealistic since users adjust their beliefs outside of Reddit and it is impossible to capture all the variance of their sympathy.
At the other extreme, $\vsym \rightarrow +\infty$ indicates that the opinion of the users is determined exclusively by factors outside of the model.
Thus, we present the results for different values of \vsym within these two extremes, setting it to $0.01$, $1$, and $100$, to account for low, medium, and high levels of individual variation.
Results do not significantly change for different values of $\vsym$.

All the priors of our model are (multivariate) Normal distributions.
We set positive mean for the priors of $N_{sub} \rightarrow P_{S}$ and $N_{sub} \rightarrow P_{L}$, under the assumption that users have a higher propensity to write in popular subreddits.
The other priors are centered in 0 and have covariance equal to the identity matrix. %

\section{Results}
\label{sec:results}

\begin{figure*}
    \centering
    \includegraphics[width=0.95\linewidth]{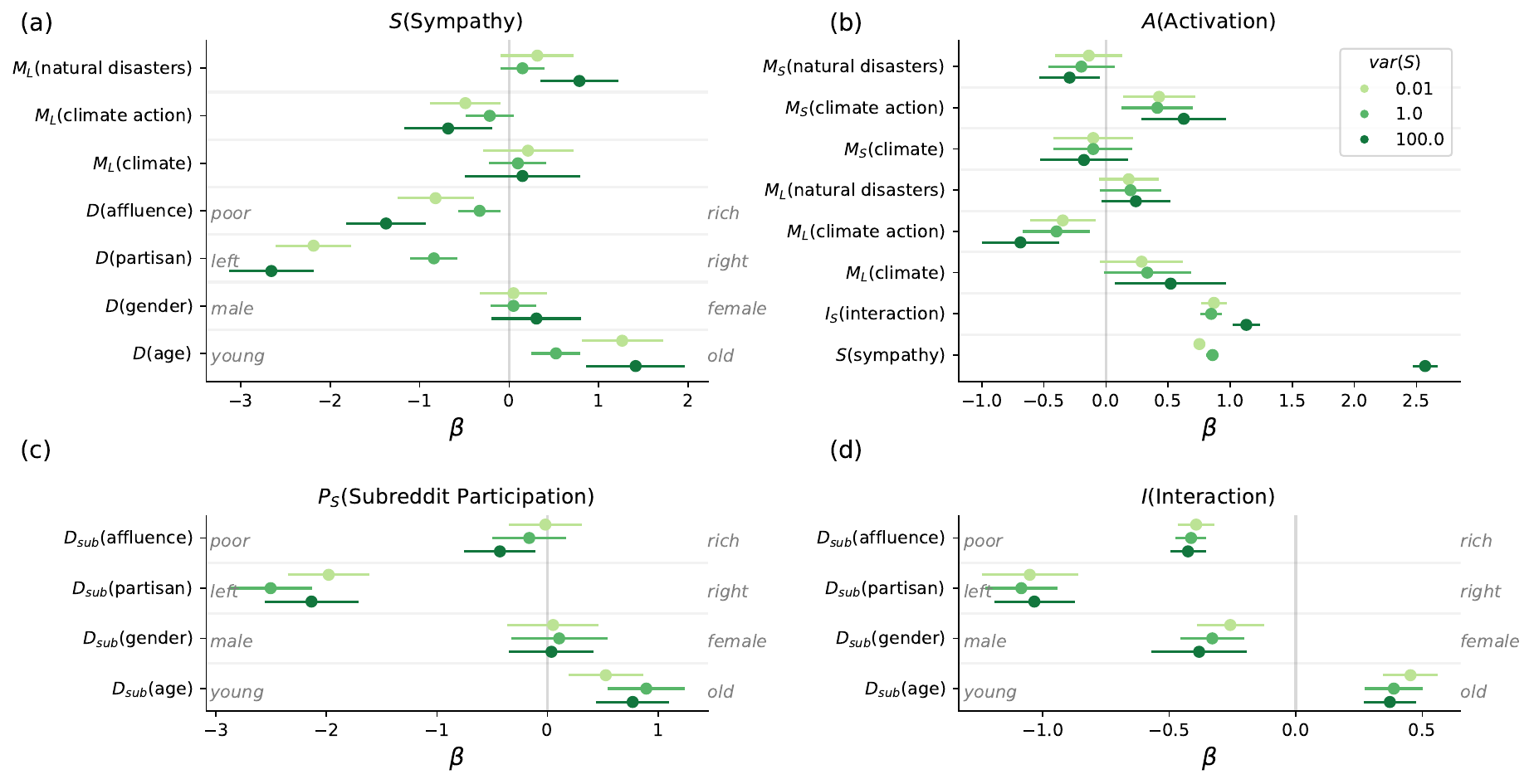}
    \caption{Mean values of the estimated parameters, with 95\% credible intervals.
    (a) The coefficients are the causal effects of the parents of \sym, \Cref{eq:sympathy}.
    These contribute to the weighted average of the mean of the normal extraction of \sym.
    (b) The coefficients are the log-odds in \Cref{eq:activation}.
    (c) The coefficients are $\log  \beta_{p1}$ in \Cref{eq:P_S}. 
    They represent the interaction term between $D_{sub}$ and \sym.
    When the coefficient is positive, an increase of \sym increases the odds of writing in a subreddit with a positive score in that sociodemographic category.
    (d) The coefficients are $\log \beta_{I1}$ in \Cref{eq:5}.
    They represent the interaction term between $P_S$ and $D_{sub}$.
    When the coefficient is positive, an increase of $P_S \cdot D_{sub}$ increases the odds of interacting with an activist.
    }
    \label{fig:coefficients_errorplot}
    \vspace{-1\baselineskip}
\end{figure*}

We now answer our research questions, using the coefficients estimated by our model.
In our Bayesian framework, in fact, each parameter provides new knowledge on the investigated process, both about the latent variables in the system and the causal paths within the model.

As introduced in \Cref{subsec:estimation}, we analyze the model under three different assumptions, ranging from a lower level of individual variation ($\vsym = 0.01$) to a medium ($\vsym = 1$) and a higher level ($\vsym = 100$).
The role of \vsym is regularization:  higher values trivially allows for a higher accuracy, as the model gets increasingly more free to align each individual \sym with \activation (see \Cref{apx:sympathy}).
For this reason, a comparison of the accuracy across these three models is meaningless, as an increase of \vsym is sufficient to increase the accuracy. 
Our study of the accuracy is limited to in-sample analysis, since an essential part of our model is the estimate of individual parameter for each user (e.g., their sociodemographic features).
Therefore, a train-test split would be unfeasible.
Ultimately, however, the goal of our model is to assess the significance of causal relations to answer our RQs, not to obtain the highest accuracy in a prediction task.

\Cref{fig:coefficients_errorplot} presents the relevant estimated parameters, grouped by the variable they affect.
We observe general robustness among the different choices of \vsym, as the coefficients describe similar causal effects, with minor differences in magnitude.
Next, we analyze these parameters to answer our research questions.\footnote{The code to analyze our data and reproduce the results is open and available at \url{https://anonymous.4open.science/r/climact-2FA4/}.}

\subsection{Media coverage (RQ1)}

For the media coverage variables, we have multiple pathways towards activation.
One is $M_L \rightarrow S \rightarrow A$.
\Cref{fig:coefficients_errorplot}b shows that the coefficient of sympathy to activation (i.e., $\beta_{A1}$ in \Cref{eq:activation}) is significantly positive for all models, as expected.
Thus, we consider the relationship $M_L \rightarrow S$, expressed by the coefficients $\beta_{S3}$ in \Cref{eq:sympathy}, depicted in \Cref{fig:coefficients_errorplot}a.
These coefficients represent how long-term media coverage affects sympathy at the end of that period.
The other pathways are the direct ones $M_S \rightarrow A$ and $M_L \rightarrow A$ ($\beta_{A3}$ and $\beta_{A4}$ in \Cref{eq:activation}, shown in \Cref{fig:coefficients_errorplot}b).
Since the media coverage variables are composed of three distinct themes, we discuss each one separately.

The broader ``Climate'' theme does not present significant effects, except in $M_L \rightarrow A$ for one model instance, thus suggesting a positive, yet mild effect of climate media coverage on activism.

Instead, ``Climate Action''---that is, the amount of media attention devoted to the climate activists actions---shows significantly positive effects in the same pathway towards \activation for all considered models (\Cref{fig:coefficients_errorplot}b).
This result suggests that media coverage of climate mobilization increases the likelihood of individuals to activate in the week following the news.
However, we also observe a negative coefficient in the long-term variables for ``Climate Action'', for two out of the three instances of the model in $M_S \rightarrow A$ (\Cref{fig:coefficients_errorplot}b), and for all instances in $M_L \rightarrow S$ (\Cref{fig:coefficients_errorplot}a).

We hypothesize that these negative coefficients are due to the bottom-up dynamics of the first wave of protests, that was successful despite not being covered by media during its build-up stage on Reddit.
In fact, two major spikes of activations are present in our dataset, the first occuring in November 2018 when \subreddit{EarthStrike} was created, and the second in September-October 2019, around the Global Climate Strike.
The first event is endogenous to Reddit and followed a long period of low media coverage on climate action (the third week of November 2018 is at the 3\textsuperscript{rd} percentile of climate action media coverage).
Conversely, the second event occurred in times of high media coverage, and this increased media attention might have influenced the engagement in climate activism.
However, this peak of {\activation}s lasted a few weeks, and the periods with the highest long-term media coverage of climate action are not associated with an increased number of {\activation}s.

Regarding the media theme of ``Natural Disasters'', \Cref{fig:coefficients_errorplot}a shows a significant postive effect in the pathway $M_L \rightarrow S$ for one of the three instances of the model.
In the same model instance, the coefficient for $M_S \rightarrow A$ is slightly negative, suggesting that awareness of climate emergencies might contribute more to the process of opinion formation on climate topics than to trigger {\activation}s directly.

In summary, media impact {\activation}s in three ways. 
Firstly, the organization of climate activism movements on Reddit was a grassroots initiative, spawned independent of media coverage.
Secondly, coverage about climate action has a positive effect on the individuals activation, with this effect decaying after about a week.
Lastly, some media coverage of extreme weather events may have a positive long-term impact on sympathy towards activism.

\subsection{Demographics (RQ2)}
From \Cref{fig:coefficients_errorplot}a, we gather that having specific sociodemographics (i.e., `left', `poor', and `old' users) increases the sympathy \sym.
Given the positive effect of \sym on \activation (\Cref{fig:coefficients_errorplot}b) these sociodemographic features impact \activation, mediated by sympathy.

\Cref{fig:coefficients_errorplot}c shows that the sympathy \sym has an impact on $P_S$, as it increases the probability of participating in `old' and `left' subreddits.
The participated subreddits also impact the probability of interacting with other activists (\Cref{fig:coefficients_errorplot}d) and the interactions have a positive effect on $A$ (\Cref{fig:coefficients_errorplot}b).
Thus, the sociodemographic features affect the activation, mediated by \sym, $P_S$ and $I$.

Finally, perhaps obviously, the user sociodemographics $D$ impact subreddit participation $P_L$, as being `left' increases the probability of participating in `left' subreddits, and analogously for all the sociodemographic dimensions.
Moreover, we estimate a positive effect of retention, as $P_L$ affects $P_S$.
As seen before, $P_S$ has an effect on $I$, that impacts $A$, so we conclude that the effect of $D$ on $A$ is also mediated by $P_L$, $P_S$, and $I$.

Note that, in the context of this study, `old' should be interpreted relative to the demographics of the platform (and similarly for the other dimensions).
In particular, considering the demographic of Reddit, an age label towards `old' is not in contrast with the Youth Climate Strikes, as 64\% of Reddit users are 18 to 29 years old~\cite{barthel2016reddit}.
A qualitative analysis of the sociodemographic scores indicates that the subreddits with lower age scores are typically about video games and entertainment, while politically and culturally engaged communities are more typical of older ages.
In particular, some popular subreddits with high correlation between $P_L$ and $A$, such as \subreddit{environment} and \subreddit{collapse}, have high scores for age.
Moreover, the 15 subreddits with the highest correlation between $P_L$ and $A$ with a sociodemographic score have a `left' partisan score, including some with an anarchist inclination such as \subreddit{COMPLETEANARCHY}, \subreddit{collapse}, and \subreddit{Anarchism} or socialist (\subreddit{ChapoTrapHouse}).

\subsection{Interactions (RQ3)}
\Cref{fig:coefficients_errorplot}b shows that the interactions positively impact \activation.
This positive effect of $I$ on $A$ can be interpreted as already net of any homophily effect due to the common sympathy, since the effect of $P_{S}$ on $I$ is controlled by \sym.
That is, given two users with the same level of sympathy who participate in similar subreddits, if one interacts with an activist, the odds of activation increase by approximately 2.7 times.
This strong positive effect can be understood in the context of the history of the Earth Strike on Reddit.
In fact, the users first joinin \subreddit{EarthStrike} migrated from other subreddits where they decided to coordinate towards global action.

\subsection{Summary of findings}

Thanks to the comprehensive causal model proposed in this work, we can finally draw comparative conclusions about the determinants of climate activism online.

For RQ1, the only consistent effect is by short-term bursts media coverage of climate action ($M_S \rightarrow A$) that decays in the long-term ($M_L \rightarrow A$).
Conversely, the path $M_L \rightarrow S \rightarrow A$ is not consistently significant.
For RQ2, the age, partisanship, and affluence of users affect $A$ via $D \rightarrow \sym \rightarrow A$.
In particular, belonging to a lower economic stratum, having left-wing political preferences, and being on the older side of Reddit's population, %
all increase the probability of \activation.
Similarly, in the path $D \rightarrow \sym \rightarrow P_S \rightarrow I \rightarrow A$, an increase in sympathy increases the likelihood of participating in subreddits with a `poor', `left', `male', and `old' user base.
These characteristics of the subreddits increase the probability of interacting with some activists, thus raising the probability of activation.
All four sociodemographic features have a significant impact on $A$ through the path $D \rightarrow P_L \rightarrow P_S \rightarrow I \rightarrow A$, as the participation in `poor', `left', `male', and `old' communities is necessarily affected by the sociodemographic features of the users.
Finally, focusing on RQ3, interactions have a strong, positive, direct effect on the activation.
That is, interacting with an activist outside of activist communities considerably increases the odds of becoming \activated.

Compared to the effects of media coverage, both sympathy for the cause and direct interactions with activists have stronger effects, with coefficients approximately twice as large.
Sympathy, in turn, is strongly affected by the sociodemographics of the user, especially their political partisanship, affluence, and age.

\begin{figure}
    \centering
    \includegraphics[width=0.9\linewidth]{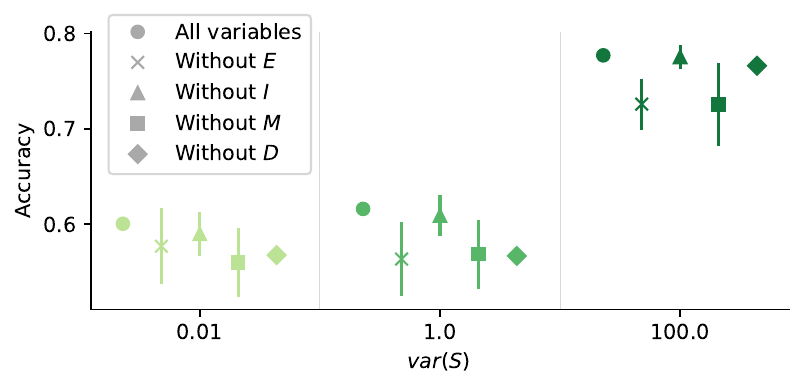}
    \caption{For each value of \vsym, we compare the distributions of accuracy of 100 samples of the predictive posterior for $A$.
    The error bars represent one standard deviation of the accuracy.
    For each \vsym, the left-hand dot refers to the experiment using all the variables of the model.
    The other dots refer to the experiment run after the removal of a variable, both long- and short-term.
    }
    \label{fig:ablation}
    \vspace{-1\baselineskip}
\end{figure}

\subsection{Ablation}
Last, we perform an ablation analysis to check the effect of neglecting some parts of the model in our results.
We run the model removing one group of variables at a time: ($i$) $E_L$ and $E_S$, ($ii$) $I$, ($iii$) $M_L$ and $M_S$ and ($iv$) $D$.
\Cref{fig:ablation} shows that the removal of the media coverage has the highest impact on the accuracy across the three models.
When removing $E$, we also note that the effects of $D_{sub}$(affluence) and $D_{sub}$(gender) switch signs.
This indicates that not controlling for user engagement might generate misleading effects on the interactions.
In particular, the full model captures that writing in `male' and `poor' subreddits increases the probability of interaction.
Instead, according to the ablated model, writing in `female' and `rich' subreddits increases it.
This effect is indeed confounded by the positive correlation of $0.38$ and $0.42$ between the engagement on Reddit and the participation in `female' and `rich' subreddits, respectively (\Cref{apx:sociodemo}), and highlights the importance of developing a comprehensive causal model.
All the other effects maintain the same signs.

We also compared our results with the ones obtained by  removing the gap between the short- and long-term windows, and obtained a correlation of 0.997 between the estimated coefficients in the two settings (\Cref{fig:temporal}).
This robustness check shows that the choice of keeping a gap period between long- and short-term past is not necessary.

\section{Discussion}
\label{sec:discussion}
We developed a rich and comprehensive causal model to study the interplay between different determinants of climate activism on Reddit.
This work represents a first attempt to apply a multi-causal model to social media data.
Our results show that the simultaneous inclusion of several causal pathways into the same model provides new insights into the underlying social process.
For instance, the interactions play a central role in the engagement in climate activism communities.
However, part of this effect is explained by the confounder of the frequented subreddits, which is controlled by the sympathy for the climate cause.
Similarly, we show that the sociodemographic features influence the \activation.
However, the ablation analysis reveals that if we omit to control for the engagement of the users, some effects change drastically and switch signs, due to the confounding effect of the correlation between engagement on Reddit and specific sociodemographic features.
Finally, we illustrate that the exclusion of media coverage from the model significantly decreases its accuracy.

The contribution of this work is two-fold.
First, we establish an operative pipeline to causally test hypotheses in the domain of computational social science using large-scale empirical data (Reddit and GDELT).
The pipeline targets the existing gap between behavioral models in social science, which are usually validated via survey or lab studies, and social media analysis, which typically draws causal conclusions only by focusing on the temporal dimension.

Second, we provide new insights into the phenomenon of climate activism online.
We unveil a temporally complex effect of media coverage of climate activism. 
On the one hand, the effect of media attention on climate action rapidly decays, with a strong short-term effect and a negative effect in the long term.
On the other hand, media coverage around climate and natural disasters is persistent over a long period.
These results hint at a trend effect for climate strikes, and a memory effect for people experiencing extreme events~\cite{torricelli2023does}.
Users participating in left-leaning and lower-income communities have a higher propensity to engage in climate activism, mediated both by their sympathy towards the cause and by the subreddits they frequent, which expose them to interactions with other activists.
We acknowledge that these sociodemographic features are an oversimplification of the users' real attributes, as they depend on the scores previously assigned to subreddits.
In the future, it would be interesting to study how different estimators of these factors affect the results.
Last, social influence has the strongest impact on user activation.

We underline that the model aims to uncover \emph{how} or the \emph{why} people engage in activism, and it is not intended to \emph{predict} user engagement.
For these reason, we do not approach the task through machine-learning, which could improve the accuracy of our estimates by leveraging spurious (i.e., non-causal) correlations in the data.
For instance, we did not measure the out-of-sample accuracy, because the latent sociodemographic features of the users are estimated during model training, and we cannot predict the \activation of a user without injecting them into the model.

We recognize that active participation in a climate activist community is not necessarily linked to joining a climate strike.
However, our framework is based on well-grounded sociological theories that apply to the online context too~\cite{van2015social}.
From the moment a user becomes active in such an activist community, they will become possible targets of mobilization by the group.
The remaining steps of the process 
happen outside of the social media sphere, 
becoming harder to study~\cite{mejova2022modeling} and
highlighting the need for broader methodologies to understand %
the full spectrum of social coordination, from digital engagement to physical mobilization and collective action.

\bibliographystyle{ACM-Reference-Format}
\bibliography{bibliography}


\begin{thebibliography}{47}

%%% the Web address.

\ifx \showCODEN    \undefined \def \showCODEN     #1{\unskip}     \fi
\ifx \showDOI      \undefined \def \showDOI       #1{#1}\fi
\ifx \showISBNx    \undefined \def \showISBNx     #1{\unskip}     \fi
\ifx \showISBNxiii \undefined \def \showISBNxiii  #1{\unskip}     \fi
\ifx \showISSN     \undefined \def \showISSN      #1{\unskip}     \fi
\ifx \showLCCN     \undefined \def \showLCCN      #1{\unskip}     \fi
\ifx \shownote     \undefined \def \shownote      #1{#1}          \fi
\ifx \showarticletitle \undefined \def \showarticletitle #1{#1}   \fi
\ifx \showURL      \undefined \def \showURL       {\relax}        \fi
% invisible to TeX
\providecommand\bibfield[2]{#2}
\providecommand\bibinfo[2]{#2}
\providecommand\natexlab[1]{#1}
\providecommand\showeprint[2][]{arXiv:#2}

\bibitem[202(2022)]%
        {2022demographics}
 \bibinfo{year}{2022}\natexlab{}.
\newblock \showarticletitle{Demographics of Emissions}.
\newblock \bibinfo{journal}{\emph{Nature Climate Change}} \bibinfo{volume}{12}, \bibinfo{number}{3} (\bibinfo{date}{March} \bibinfo{year}{2022}), \bibinfo{pages}{209--209}.
\newblock
\showISSN{1758-6798}
\urldef\tempurl%
\url{https://doi.org/10.1038/s41558-022-01325-5}
\showDOI{\tempurl}


\bibitem[Balsamo et~al\mbox{.}(2019)]%
        {balsamo2019firsthand}
\bibfield{author}{\bibinfo{person}{Duilio Balsamo}, \bibinfo{person}{Paolo Bajardi}, {and} \bibinfo{person}{Andr{\'e} Panisson}.} \bibinfo{year}{2019}\natexlab{}.
\newblock \showarticletitle{Firsthand opiates abuse on social media: monitoring geospatial patterns of interest through a digital cohort}. In \bibinfo{booktitle}{\emph{The World Wide Web Conference}}. \bibinfo{pages}{2572--2579}.
\newblock


\bibitem[Bamberg et~al\mbox{.}(2015)]%
        {bamberg2015collective}
\bibfield{author}{\bibinfo{person}{Sebastian Bamberg}, \bibinfo{person}{Jonas Rees}, {and} \bibinfo{person}{Sebastian Seebauer}.} \bibinfo{year}{2015}\natexlab{}.
\newblock \showarticletitle{Collective climate action: Determinants of participation intention in community-based pro-environmental initiatives}.
\newblock \bibinfo{journal}{\emph{Journal of Environmental Psychology}}  \bibinfo{volume}{43} (\bibinfo{year}{2015}), \bibinfo{pages}{155--165}.
\newblock


\bibitem[Barthel et~al\mbox{.}(2016)]%
        {barthel2016reddit}
\bibfield{author}{\bibinfo{person}{Michael Barthel}, \bibinfo{person}{Galen Stocking}, \bibinfo{person}{Jess Holcomb}, {and} \bibinfo{person}{Amy Mitchell}.} \bibinfo{year}{2016}\natexlab{}.
\newblock \showarticletitle{Reddit news users more likely to be male, young and digital in their news preferences}.
\newblock \bibinfo{journal}{\emph{Pew Research Center}} (\bibinfo{year}{2016}).
\newblock


\bibitem[Bliuc et~al\mbox{.}(2007)]%
        {bliuc2007opinion}
\bibfield{author}{\bibinfo{person}{Ana-Maria Bliuc}, \bibinfo{person}{Craig McGarty}, \bibinfo{person}{Katherine Reynolds}, {and} \bibinfo{person}{Daniela Muntele}.} \bibinfo{year}{2007}\natexlab{}.
\newblock \showarticletitle{Opinion-based group membership as a predictor of commitment to political action}.
\newblock \bibinfo{journal}{\emph{European journal of social psychology}} \bibinfo{volume}{37}, \bibinfo{number}{1} (\bibinfo{year}{2007}), \bibinfo{pages}{19--32}.
\newblock


\bibitem[Chen et~al\mbox{.}(2023)]%
        {chen2023climate}
\bibfield{author}{\bibinfo{person}{Kaiping Chen}, \bibinfo{person}{Amanda~L Molder}, \bibinfo{person}{Zening Duan}, \bibinfo{person}{Shelley Boulianne}, \bibinfo{person}{Christopher Eckart}, \bibinfo{person}{Prince Mallari}, {and} \bibinfo{person}{Diyi Yang}.} \bibinfo{year}{2023}\natexlab{}.
\newblock \showarticletitle{How climate movement actors and news media frame climate change and strike: Evidence from analyzing twitter and news media discourse from 2018 to 2021}.
\newblock \bibinfo{journal}{\emph{The International Journal of Press/Politics}} \bibinfo{volume}{28}, \bibinfo{number}{2} (\bibinfo{year}{2023}), \bibinfo{pages}{384--413}.
\newblock


\bibitem[Doell et~al\mbox{.}(2024)]%
        {doell2024international}
\bibfield{author}{\bibinfo{person}{Kimberly~C Doell}, \bibinfo{person}{Boryana Todorova}, \bibinfo{person}{Madalina Vlasceanu}, \bibinfo{person}{JB Bak-Coleman}, \bibinfo{person}{Jay~J Van~Bavel}, {and} \bibinfo{person}{Ekaterina Pronizius}.} \bibinfo{year}{2024}\natexlab{}.
\newblock \showarticletitle{The International Climate Psychology Collaboration: Climate change-related data collected from 63 countries}.
\newblock  (\bibinfo{year}{2024}).
\newblock


\bibitem[Edenhofer(2015)]%
        {ipcc2014}
\bibfield{author}{\bibinfo{person}{Ottmar Edenhofer}.} \bibinfo{year}{2015}\natexlab{}.
\newblock \bibinfo{booktitle}{\emph{Climate change 2014: mitigation of climate change}}. Vol.~\bibinfo{volume}{3}.
\newblock \bibinfo{publisher}{Cambridge University Press}.
\newblock
\urldef\tempurl%
\url{https://archive.ipcc.ch/pdf/assessment-report/ar5/wg3/drafts/fgd/ipcc_wg3_ar5_summary-for-policymakers_may-version.pdf}
\showURL{%
\tempurl}


\bibitem[Emmerling et~al\mbox{.}(2024)]%
        {emmerling2024multi}
\bibfield{author}{\bibinfo{person}{Johannes Emmerling}, \bibinfo{person}{Pietro Andreoni}, \bibinfo{person}{Ioannis Charalampidis}, \bibinfo{person}{Shouro Dasgupta}, \bibinfo{person}{Francis Dennig}, \bibinfo{person}{Toon Vandyck}, \bibinfo{person}{Simon Feindt}, \bibinfo{person}{Dimitris Fragkiadakis}, \bibinfo{person}{Panagiotis Fragkos}, \bibinfo{person}{Shinichiro Fujimori}, {et~al\mbox{.}}} \bibinfo{year}{2024}\natexlab{}.
\newblock \showarticletitle{A Multi-Model Assessment of Inequality and Climate Change}.
\newblock  (\bibinfo{year}{2024}).
\newblock


\bibitem[Falkenberg et~al\mbox{.}(2022)]%
        {falkenberg2022growing}
\bibfield{author}{\bibinfo{person}{Max Falkenberg}, \bibinfo{person}{Alessandro Galeazzi}, \bibinfo{person}{Maddalena Torricelli}, \bibinfo{person}{Niccol{\`o} Di~Marco}, \bibinfo{person}{Francesca Larosa}, \bibinfo{person}{Madalina Sas}, \bibinfo{person}{Amin Mekacher}, \bibinfo{person}{Warren Pearce}, \bibinfo{person}{Fabiana Zollo}, \bibinfo{person}{Walter Quattrociocchi}, {et~al\mbox{.}}} \bibinfo{year}{2022}\natexlab{}.
\newblock \showarticletitle{Growing polarization around climate change on social media}.
\newblock \bibinfo{journal}{\emph{Nature Climate Change}} \bibinfo{volume}{12}, \bibinfo{number}{12} (\bibinfo{year}{2022}), \bibinfo{pages}{1114--1121}.
\newblock


\bibitem[Friedkin(1998)]%
        {friedkin1998structural}
\bibfield{author}{\bibinfo{person}{Noah~E Friedkin}.} \bibinfo{year}{1998}\natexlab{}.
\newblock \bibinfo{booktitle}{\emph{A structural theory of social influence.}}
\newblock \bibinfo{publisher}{Cambridge University Press}.
\newblock


\bibitem[Friedkin and Johnsen(1990)]%
        {friedkin1990social}
\bibfield{author}{\bibinfo{person}{Noah~E Friedkin} {and} \bibinfo{person}{Eugene~C Johnsen}.} \bibinfo{year}{1990}\natexlab{}.
\newblock \showarticletitle{Social influence and opinions}.
\newblock \bibinfo{journal}{\emph{Journal of Mathematical Sociology}} \bibinfo{volume}{15}, \bibinfo{number}{3-4} (\bibinfo{year}{1990}), \bibinfo{pages}{193--206}.
\newblock


\bibitem[Gaupp and Eker(2024)]%
        {gaupp2024climate}
\bibfield{author}{\bibinfo{person}{Franziska Gaupp} {and} \bibinfo{person}{Sibel Eker}.} \bibinfo{year}{2024}\natexlab{}.
\newblock \showarticletitle{Climate Activism, Social Media and Behavioural Change: A Literature Review}.
\newblock  (\bibinfo{year}{2024}).
\newblock


\bibitem[Han and Ahn(2020)]%
        {han2020youth}
\bibfield{author}{\bibinfo{person}{Heejin Han} {and} \bibinfo{person}{Sang~Wuk Ahn}.} \bibinfo{year}{2020}\natexlab{}.
\newblock \showarticletitle{Youth mobilization to stop global climate change: Narratives and impact}.
\newblock \bibinfo{journal}{\emph{Sustainability}} \bibinfo{volume}{12}, \bibinfo{number}{10} (\bibinfo{year}{2020}), \bibinfo{pages}{4127}.
\newblock


\bibitem[Hoffman et~al\mbox{.}(2013)]%
        {hoffman2013stochastic}
\bibfield{author}{\bibinfo{person}{Matthew~D Hoffman}, \bibinfo{person}{David~M Blei}, \bibinfo{person}{Chong Wang}, {and} \bibinfo{person}{John Paisley}.} \bibinfo{year}{2013}\natexlab{}.
\newblock \showarticletitle{Stochastic variational inference}.
\newblock \bibinfo{journal}{\emph{Journal of Machine Learning Research}} (\bibinfo{year}{2013}).
\newblock


\bibitem[IPCC(2022)]%
        {ipcc2022}
\bibfield{author}{\bibinfo{person}{IPCC}.} \bibinfo{year}{2022}\natexlab{}.
\newblock \bibinfo{booktitle}{\emph{Summary for Policymakers}}.
\newblock \bibinfo{publisher}{Cambridge University Press}, \bibinfo{address}{Cambridge, UK}, \bibinfo{pages}{In Press}.
\newblock
\urldef\tempurl%
\url{https://www.ipcc.ch/report/sixth-assessment-report-working-group-ii}
\showURL{%
\tempurl}
\newblock
\shownote{In Press}.


\bibitem[Islam et~al\mbox{.}(2013)]%
        {islam2013investigation}
\bibfield{author}{\bibinfo{person}{Md~Mofakkarul Islam}, \bibinfo{person}{Andrew Barnes}, {and} \bibinfo{person}{Luiza Toma}.} \bibinfo{year}{2013}\natexlab{}.
\newblock \showarticletitle{An investigation into climate change scepticism among farmers}.
\newblock \bibinfo{journal}{\emph{Journal of Environmental Psychology}}  \bibinfo{volume}{34} (\bibinfo{year}{2013}), \bibinfo{pages}{137--150}.
\newblock


\bibitem[Kirilenko et~al\mbox{.}(2015)]%
        {kirilenko2015people}
\bibfield{author}{\bibinfo{person}{Andrei~P Kirilenko}, \bibinfo{person}{Tatiana Molodtsova}, {and} \bibinfo{person}{Svetlana~O Stepchenkova}.} \bibinfo{year}{2015}\natexlab{}.
\newblock \showarticletitle{People as sensors: Mass media and local temperature influence climate change discussion on Twitter}.
\newblock \bibinfo{journal}{\emph{Global Environmental Change}}  \bibinfo{volume}{30} (\bibinfo{year}{2015}), \bibinfo{pages}{92--100}.
\newblock


\bibitem[Lenti et~al\mbox{.}(2024a)]%
        {lenti2024likelihood}
\bibfield{author}{\bibinfo{person}{Jacopo Lenti}, \bibinfo{person}{Corrado Monti}, {and} \bibinfo{person}{Gianmarco De~Francisci~Morales}.} \bibinfo{year}{2024}\natexlab{a}.
\newblock \showarticletitle{Likelihood-Based Methods Improve Parameter Estimation in Opinion Dynamics Models}.
\newblock \bibinfo{journal}{\emph{Proceedings of the 17th ACM International Conference on Web Search and Data Mining}} (\bibinfo{year}{2024}).
\newblock


\bibitem[Lenti et~al\mbox{.}(2024b)]%
        {lenti2024variational}
\bibfield{author}{\bibinfo{person}{Jacopo Lenti}, \bibinfo{person}{Fabrizio Silvestri}, {and} \bibinfo{person}{Gianmarco De~Francisci Morales}.} \bibinfo{year}{2024}\natexlab{b}.
\newblock \showarticletitle{Variational Inference of Parameters in Opinion Dynamics Models}.
\newblock \bibinfo{journal}{\emph{arXiv preprint arXiv:2403.05358}} (\bibinfo{year}{2024}).
\newblock


\bibitem[Massachs et~al\mbox{.}(2020)]%
        {massachs2020roots}
\bibfield{author}{\bibinfo{person}{Joan Massachs}, \bibinfo{person}{Corrado Monti}, \bibinfo{person}{Gianmarco De~Francisci Morales}, {and} \bibinfo{person}{Francesco Bonchi}.} \bibinfo{year}{2020}\natexlab{}.
\newblock \showarticletitle{Roots of trumpism: Homophily and social feedback in donald trump support on reddit}. In \bibinfo{booktitle}{\emph{Proceedings of the 12th ACM conference on Web Science}}. \bibinfo{pages}{49--58}.
\newblock


\bibitem[Mavrodieva et~al\mbox{.}(2019)]%
        {mavrodieva2019role}
\bibfield{author}{\bibinfo{person}{AV Mavrodieva}, \bibinfo{person}{OK Rachman}, \bibinfo{person}{VB Harahap}, {and} \bibinfo{person}{R Shaw}.} \bibinfo{year}{2019}\natexlab{}.
\newblock \bibinfo{title}{Role of social media as a soft power tool in raising public awareness and engagement in addressing climate change. Climate, 7 (10), 122}.
\newblock
\newblock


\bibitem[McGarty et~al\mbox{.}(2009)]%
        {mcgarty2009collective}
\bibfield{author}{\bibinfo{person}{Craig McGarty}, \bibinfo{person}{Ana-Maria Bliuc}, \bibinfo{person}{Emma~F Thomas}, {and} \bibinfo{person}{Renata Bongiorno}.} \bibinfo{year}{2009}\natexlab{}.
\newblock \showarticletitle{Collective action as the material expression of opinion-based group membership}.
\newblock \bibinfo{journal}{\emph{Journal of Social Issues}} \bibinfo{volume}{65}, \bibinfo{number}{4} (\bibinfo{year}{2009}), \bibinfo{pages}{839--857}.
\newblock


\bibitem[Mejova et~al\mbox{.}(2022)]%
        {mejova2022modeling}
\bibfield{author}{\bibinfo{person}{Yelena Mejova}, \bibinfo{person}{Jisun An}, \bibinfo{person}{Gianmarco De~Francisci Morales}, {and} \bibinfo{person}{Haewoon Kwak}.} \bibinfo{year}{2022}\natexlab{}.
\newblock \showarticletitle{Modeling {Political} {Activism} around {Gun} {Debate} via {Social} {Media}}.
\newblock \bibinfo{journal}{\emph{ACM Transactions on Social Computing}} (\bibinfo{date}{April} \bibinfo{year}{2022}), \bibinfo{pages}{3532102}.
\newblock
\showISSN{2469-7818, 2469-7826}
\urldef\tempurl%
\url{https://doi.org/10.1145/3532102}
\showDOI{\tempurl}


\bibitem[Mejova and Kourtellis(2021)]%
        {mejova2021youtubing}
\bibfield{author}{\bibinfo{person}{Yelena Mejova} {and} \bibinfo{person}{Nicolas Kourtellis}.} \bibinfo{year}{2021}\natexlab{}.
\newblock \showarticletitle{Youtubing at home: Media sharing behavior change as proxy for mobility around covid-19 lockdowns}. In \bibinfo{booktitle}{\emph{Proceedings of the 13th ACM Web Science Conference 2021}}. \bibinfo{pages}{272--281}.
\newblock


\bibitem[Minici et~al\mbox{.}(2022)]%
        {minici2022cascade}
\bibfield{author}{\bibinfo{person}{Marco Minici}, \bibinfo{person}{Federico Cinus}, \bibinfo{person}{Corrado Monti}, \bibinfo{person}{Francesco Bonchi}, {and} \bibinfo{person}{Giuseppe Manco}.} \bibinfo{year}{2022}\natexlab{}.
\newblock \showarticletitle{Cascade-based echo chamber detection}. In \bibinfo{booktitle}{\emph{Proceedings of the 31st ACM International Conference on Information \& Knowledge Management}}. \bibinfo{pages}{1511--1520}.
\newblock


\bibitem[Monti et~al\mbox{.}(2020)]%
        {monti2020learning}
\bibfield{author}{\bibinfo{person}{Corrado Monti}, \bibinfo{person}{Gianmarco De~Francisci~Morales}, {and} \bibinfo{person}{Francesco Bonchi}.} \bibinfo{year}{2020}\natexlab{}.
\newblock \showarticletitle{Learning opinion dynamics from social traces}. In \bibinfo{booktitle}{\emph{Proceedings of the 26th ACM SIGKDD International Conference on Knowledge Discovery \& Data Mining}}. \bibinfo{pages}{764--773}.
\newblock


\bibitem[Monti et~al\mbox{.}(2021)]%
        {monti2021learning}
\bibfield{author}{\bibinfo{person}{Corrado Monti}, \bibinfo{person}{Giuseppe Manco}, \bibinfo{person}{Cigdem Aslay}, {and} \bibinfo{person}{Francesco Bonchi}.} \bibinfo{year}{2021}\natexlab{}.
\newblock \showarticletitle{Learning ideological embeddings from information cascades}. In \bibinfo{booktitle}{\emph{Proceedings of the 30th ACM International Conference on Information \& Knowledge Management}}. \bibinfo{pages}{1325--1334}.
\newblock


\bibitem[Neas et~al\mbox{.}(2022)]%
        {neas2022young}
\bibfield{author}{\bibinfo{person}{Sally Neas}, \bibinfo{person}{Ann Ward}, {and} \bibinfo{person}{Benjamin Bowman}.} \bibinfo{year}{2022}\natexlab{}.
\newblock \showarticletitle{Young people's climate activism: A review of the literature}.
\newblock \bibinfo{journal}{\emph{Frontiers in Political Science}}  \bibinfo{volume}{4} (\bibinfo{year}{2022}), \bibinfo{pages}{940876}.
\newblock


\bibitem[Newman and Noy(2023)]%
        {newman2023global}
\bibfield{author}{\bibinfo{person}{Rebecca Newman} {and} \bibinfo{person}{Ilan Noy}.} \bibinfo{year}{2023}\natexlab{}.
\newblock \showarticletitle{The global costs of extreme weather that are attributable to climate change}.
\newblock \bibinfo{journal}{\emph{Nature Communications}} \bibinfo{volume}{14}, \bibinfo{number}{1} (\bibinfo{year}{2023}), \bibinfo{pages}{6103}.
\newblock


\bibitem[Pearl(2009)]%
        {pearl2009causality}
\bibfield{author}{\bibinfo{person}{Judea Pearl}.} \bibinfo{year}{2009}\natexlab{}.
\newblock \bibinfo{booktitle}{\emph{Causality}}.
\newblock \bibinfo{publisher}{Cambridge university press}.
\newblock


\bibitem[Phadke et~al\mbox{.}(2021)]%
        {phadke2021makes}
\bibfield{author}{\bibinfo{person}{Shruti Phadke}, \bibinfo{person}{Mattia Samory}, {and} \bibinfo{person}{Tanushree Mitra}.} \bibinfo{year}{2021}\natexlab{}.
\newblock \showarticletitle{What makes people join conspiracy communities? role of social factors in conspiracy engagement}.
\newblock \bibinfo{journal}{\emph{Proceedings of the ACM on Human-Computer Interaction}} \bibinfo{volume}{4}, \bibinfo{number}{CSCW3} (\bibinfo{year}{2021}), \bibinfo{pages}{1--30}.
\newblock


\bibitem[Purohit et~al\mbox{.}(2011)]%
        {purohit2011understanding}
\bibfield{author}{\bibinfo{person}{Hemant Purohit}, \bibinfo{person}{Yiye Ruan}, \bibinfo{person}{Amruta Joshi}, \bibinfo{person}{Srinivasan Parthasarathy}, {and} \bibinfo{person}{Amit Sheth}.} \bibinfo{year}{2011}\natexlab{}.
\newblock \showarticletitle{Understanding user-community engagement by multi-faceted features: A case study on twitter}. In \bibinfo{booktitle}{\emph{WWW 2011 Workshop on Social Media Engagement (SoME)}}.
\newblock


\bibitem[Roxburgh et~al\mbox{.}(2019)]%
        {roxburgh2019characterising}
\bibfield{author}{\bibinfo{person}{Nicholas Roxburgh}, \bibinfo{person}{Dabo Guan}, \bibinfo{person}{Kong~Joo Shin}, \bibinfo{person}{William Rand}, \bibinfo{person}{Shunsuke Managi}, \bibinfo{person}{Robin Lovelace}, {and} \bibinfo{person}{Jing Meng}.} \bibinfo{year}{2019}\natexlab{}.
\newblock \showarticletitle{Characterising climate change discourse on social media during extreme weather events}.
\newblock \bibinfo{journal}{\emph{Global environmental change}}  \bibinfo{volume}{54} (\bibinfo{year}{2019}), \bibinfo{pages}{50--60}.
\newblock


\bibitem[Russo et~al\mbox{.}(2023)]%
        {russo2023stranger}
\bibfield{author}{\bibinfo{person}{Giuseppe Russo}, \bibinfo{person}{Manoel~Horta Ribeiro}, {and} \bibinfo{person}{Robert West}.} \bibinfo{year}{2023}\natexlab{}.
\newblock \showarticletitle{Stranger Danger! Cross-Community Interactions with Fringe Users Increase the Growth of Fringe Communities on Reddit}.
\newblock \bibinfo{journal}{\emph{arXiv preprint arXiv:2310.12186}} (\bibinfo{year}{2023}).
\newblock


\bibitem[Sabherwal et~al\mbox{.}(2021)]%
        {sabherwal2021greta}
\bibfield{author}{\bibinfo{person}{Anandita Sabherwal}, \bibinfo{person}{Matthew~T Ballew}, \bibinfo{person}{Sander van Der~Linden}, \bibinfo{person}{Abel Gustafson}, \bibinfo{person}{Matthew~H Goldberg}, \bibinfo{person}{Edward~W Maibach}, \bibinfo{person}{John~E Kotcher}, \bibinfo{person}{Janet~K Swim}, \bibinfo{person}{Seth~A Rosenthal}, {and} \bibinfo{person}{Anthony Leiserowitz}.} \bibinfo{year}{2021}\natexlab{}.
\newblock \showarticletitle{The Greta Thunberg Effect: Familiarity with Greta Thunberg predicts intentions to engage in climate activism in the United States}.
\newblock \bibinfo{journal}{\emph{Journal of applied social psychology}} \bibinfo{volume}{51}, \bibinfo{number}{4} (\bibinfo{year}{2021}), \bibinfo{pages}{321--333}.
\newblock


\bibitem[Schulte et~al\mbox{.}(2020)]%
        {schulte2020social}
\bibfield{author}{\bibinfo{person}{Maxie Schulte}, \bibinfo{person}{Sebastian Bamberg}, \bibinfo{person}{Jonas Rees}, {and} \bibinfo{person}{Philipp Rollin}.} \bibinfo{year}{2020}\natexlab{}.
\newblock \showarticletitle{Social identity as a key concept for connecting transformative societal change with individual environmental activism}.
\newblock \bibinfo{journal}{\emph{Journal of Environmental Psychology}}  \bibinfo{volume}{72} (\bibinfo{year}{2020}), \bibinfo{pages}{101525}.
\newblock


\bibitem[Sisco et~al\mbox{.}(2017)]%
        {sisco2017extreme}
\bibfield{author}{\bibinfo{person}{Matthew~R Sisco}, \bibinfo{person}{Valentina Bosetti}, {and} \bibinfo{person}{Elke~U Weber}.} \bibinfo{year}{2017}\natexlab{}.
\newblock \showarticletitle{When do extreme weather events generate attention to climate change?}
\newblock \bibinfo{journal}{\emph{Climatic change}}  \bibinfo{volume}{143} (\bibinfo{year}{2017}), \bibinfo{pages}{227--241}.
\newblock


\bibitem[Tajfel(1978)]%
        {tajfel1978social}
\bibfield{author}{\bibinfo{person}{H Tajfel}.} \bibinfo{year}{1978}\natexlab{}.
\newblock \showarticletitle{Social categorization, social identity and social comparison}.
\newblock \bibinfo{journal}{\emph{Differentiation between Social Groups: Studies in the Social Psychology of Intergroup Relations}} (\bibinfo{year}{1978}).
\newblock


\bibitem[Tajfel(1979)]%
        {tajfel1979integrative}
\bibfield{author}{\bibinfo{person}{Henri Tajfel}.} \bibinfo{year}{1979}\natexlab{}.
\newblock \showarticletitle{An integrative theory of intergroup conflict}.
\newblock \bibinfo{journal}{\emph{Intergroup relations: Essential readings}} (\bibinfo{year}{1979}).
\newblock


\bibitem[Torricelli et~al\mbox{.}(2023)]%
        {torricelli2023does}
\bibfield{author}{\bibinfo{person}{Maddalena Torricelli}, \bibinfo{person}{Max Falkenberg}, \bibinfo{person}{Alessandro Galeazzi}, \bibinfo{person}{Fabiana Zollo}, \bibinfo{person}{Walter Quattrociocchi}, {and} \bibinfo{person}{Andrea Baronchelli}.} \bibinfo{year}{2023}\natexlab{}.
\newblock \showarticletitle{How does extreme weather impact the climate change discourse? Insights from the Twitter discussion on hurricanes}.
\newblock \bibinfo{journal}{\emph{Plos Climate}} \bibinfo{volume}{2}, \bibinfo{number}{11} (\bibinfo{year}{2023}), \bibinfo{pages}{e0000277}.
\newblock


\bibitem[Van~der Linden(2015)]%
        {van2015social}
\bibfield{author}{\bibinfo{person}{Sander Van~der Linden}.} \bibinfo{year}{2015}\natexlab{}.
\newblock \showarticletitle{The social-psychological determinants of climate change risk perceptions: Towards a comprehensive model}.
\newblock \bibinfo{journal}{\emph{Journal of environmental psychology}}  \bibinfo{volume}{41} (\bibinfo{year}{2015}), \bibinfo{pages}{112--124}.
\newblock


\bibitem[Van~Stekelenburg and Klandermans(2013)]%
        {van2013social}
\bibfield{author}{\bibinfo{person}{Jacquelien Van~Stekelenburg} {and} \bibinfo{person}{Bert Klandermans}.} \bibinfo{year}{2013}\natexlab{}.
\newblock \showarticletitle{The social psychology of protest}.
\newblock \bibinfo{journal}{\emph{Current Sociology}} \bibinfo{volume}{61}, \bibinfo{number}{5-6} (\bibinfo{year}{2013}), \bibinfo{pages}{886--905}.
\newblock


\bibitem[Van~Zomeren and Iyer(2009)]%
        {van2009introduction}
\bibfield{author}{\bibinfo{person}{Martijn Van~Zomeren} {and} \bibinfo{person}{Aarti Iyer}.} \bibinfo{year}{2009}\natexlab{}.
\newblock \bibinfo{title}{Introduction to the social and psychological dynamics of collective action}.
\newblock , \bibinfo{numpages}{645--660}~pages.
\newblock


\bibitem[Waller and Anderson(2021)]%
        {waller2021quantifying}
\bibfield{author}{\bibinfo{person}{Isaac Waller} {and} \bibinfo{person}{Ashton Anderson}.} \bibinfo{year}{2021}\natexlab{}.
\newblock \showarticletitle{Quantifying social organization and political polarization in online platforms}.
\newblock \bibinfo{journal}{\emph{Nature}} \bibinfo{volume}{600}, \bibinfo{number}{7888} (\bibinfo{year}{2021}), \bibinfo{pages}{264--268}.
\newblock


\bibitem[Wang and Kim(2018)]%
        {wang2018analysis}
\bibfield{author}{\bibinfo{person}{Jaesun Wang} {and} \bibinfo{person}{Seoyong Kim}.} \bibinfo{year}{2018}\natexlab{}.
\newblock \showarticletitle{Analysis of the impact of values and perception on climate change skepticism and its implication for public policy}.
\newblock \bibinfo{journal}{\emph{Climate}} \bibinfo{volume}{6}, \bibinfo{number}{4} (\bibinfo{year}{2018}), \bibinfo{pages}{99}.
\newblock


\bibitem[Zhang et~al\mbox{.}(2017)]%
        {zhang2017community}
\bibfield{author}{\bibinfo{person}{Justine Zhang}, \bibinfo{person}{William Hamilton}, \bibinfo{person}{Cristian Danescu-Niculescu-Mizil}, \bibinfo{person}{Dan Jurafsky}, {and} \bibinfo{person}{Jure Leskovec}.} \bibinfo{year}{2017}\natexlab{}.
\newblock \showarticletitle{Community identity and user engagement in a multi-community landscape}. In \bibinfo{booktitle}{\emph{Proceedings of the international AAAI conference on web and social media}}, Vol.~\bibinfo{volume}{11}. \bibinfo{pages}{377--386}.
\newblock


\end{thebibliography}

\appendix

\renewcommand\thefigure{\thesection.\arabic{figure}}
\renewcommand\thetable{\thesection.\arabic{table}}
\setcounter{figure}{0}
\setcounter{table}{0}

\FloatBarrier
\section{Subreddits}
\label{apx:subreddits}

To capture the climate activism movements, we consider all the organizations that endorsed Earth Strike\footnote{\url{https://en.wikipedia.org/wiki/Earth_Strike}}.
We select the organizations with a subreddit with the same name, and we focus only on the subreddits that explicitly mention climate action in their bio (or in their official Webpage if the bio is missing).

This is our final list of subreddits, and the relative bio (as of 2024/07/10):
\begin{squishlist}
\item \subreddit{FridaysForFuture}: ``A sub dedicated to the international movement of students who skip class on Fridays to demand action to prevent climate change.''
\item \subreddit{SunriseMovement}: ``Sunrise is a movement to stop climate change and create millions of good jobs in the process.''
\item \subreddit{ExtinctionRebellion}: ``This is our darkest hour. Humanity finds itself embroiled in an event unprecedented in its history, one which, unless immediately addressed, will catapult us further into the destruction of all we hold dear: our nations, its peoples, our ecosystems, and the future of generations to come. The science is clear: we are in the midst of the sixth mass extinction event on planet Earth and we will face catastrophe if we do not act swiftly and robustly[…] The Extinction Rebellion is a lawfully abiding direct action movement challenging inaction over dangerous climate change.''
\item \subreddit{EarthStrike}: ``Earth Strike is a grassroots labor-environmental movement focused on organizing a GLOBAL GENERAL STRIKE TO SAVE THE PLANET! [\ldots] This subreddit is for discussion about Earth Strike, a general strike movement for climate action.''
\item \subreddit{ClimateRealityProject}: ``Our mission is to catalyze a global solution to the climate crisis by making urgent action a necessity across every level of society.''
\item \subreddit{350ppm}: ``A subreddit for the organizing of the Climate Action Movement and various Campaigns.''
\item \subreddit{sierraclub}: [from the webpage] ``Sierra Club is working tirelessly to protect wildlife and wild places, ensure clean air and water for all, and fight the devastating effects of climate change.. Sierra Club is the most historic grassroots environmental organization in the country. For more than 130 years, we’ve gathered millions of activists and volunteers to fight for the places, people, and planet we all love.''
\end{squishlist}

We excluded several subreddits that endorsed Earth Strike but that are not strictly related to climate action, and instead usually call for a more generic environmental or social action: \subreddit{Amnesty}, \subreddit{greenpeace}, \subreddit{AmnestyInternational}, \subreddit{OxfamAmerica}, \subreddit{WWF}, \subreddit{LushCosmetics}, \subreddit{198Methods},  and \subreddit{MarchForScience}.

\begin{figure}[th!]
    \centering
    \includegraphics[width=0.8\linewidth]{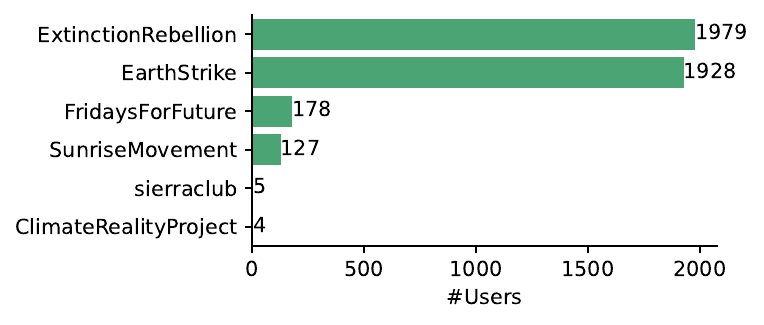}
    \caption{Number of activated users per subreddit in our final user selection. %
    }
    \label{fig:subreddits}
\end{figure}

\FloatBarrier
\section{Robustness}
\label{apx:robustness}

\begin{figure}[th!]
    \centering
    \includegraphics[width=0.6\linewidth]{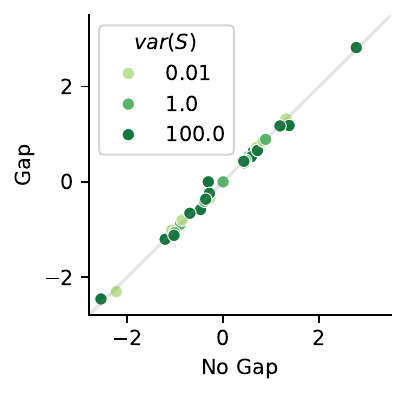}
    \caption{
    Each dot represents a parameter or latent variable of the model, excluding $D$ and \sym.
    We compare the estimates of the mean of these parameters, with and without the gap between long- and short-term periods.
    Given the correlation of 99.7\%, the choice of the endpoint of the long-term period between $t_{A-W}$ and $t_{A - 5W}$ does not affect the results.}
    \label{fig:robustness}
\end{figure}

\FloatBarrier
\section{Georeference}
\label{apx:georef}

\begin{figure}[th!]
    \centering
    \includegraphics[width=0.6\linewidth]{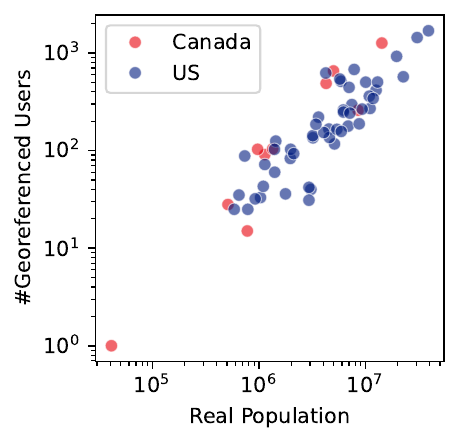}
    \caption{
    Each point represents a US State (blue) or a Canadian province (red).
    The x-coordinates are the real population of the administrative area.
    The y-coordinates are the number of georeferences per administrative area obtained in the sample of 17k users from our dataset.
    }
    \label{fig:georeference}
\end{figure}

\clearpage

\FloatBarrier
\section{Sympathy}
\label{apx:sympathy}

\begin{figure}[th!]
    \centering
    \includegraphics[width=0.9\linewidth]{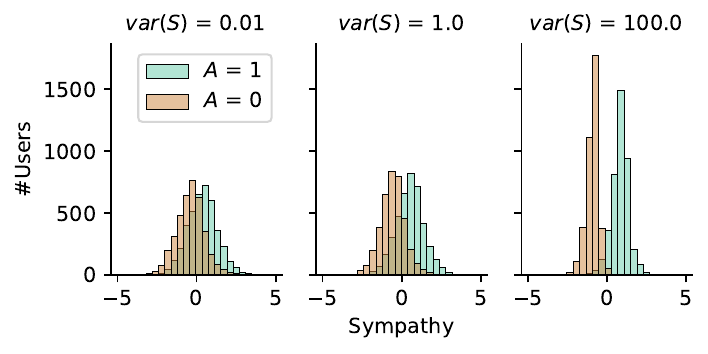}
    \caption{The plots show the distributions of \sym at the varying of \vsym.
    The separation of the distributions increases with \vsym, which reflects the increased alignment with $A$.
    }
    \label{fig:hist_sympathy}
\end{figure}

\FloatBarrier
\section{Sociodemographic Features}
\label{apx:sociodemo}

\begin{figure}[th!]
\centering

\subfloat{\includegraphics[width=0.2\textwidth]{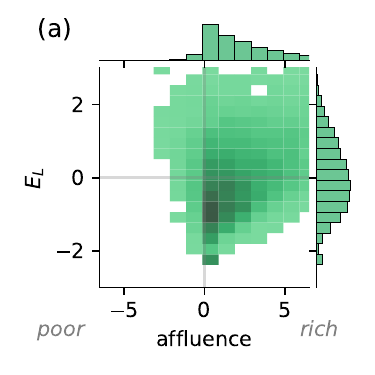}} 
\qquad
\subfloat{\includegraphics[width=0.2\textwidth]{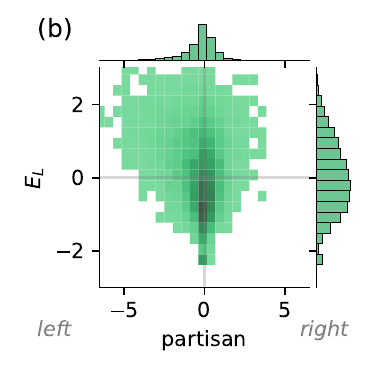}}
\qquad
\subfloat{\includegraphics[width=0.2\textwidth]{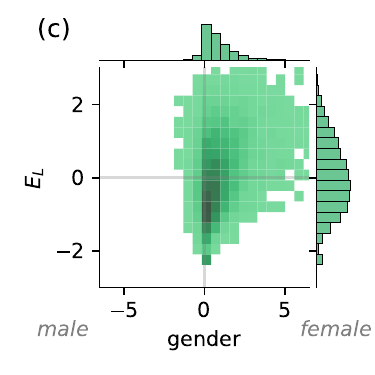}}
\qquad
\subfloat{\includegraphics[width=0.2\textwidth]{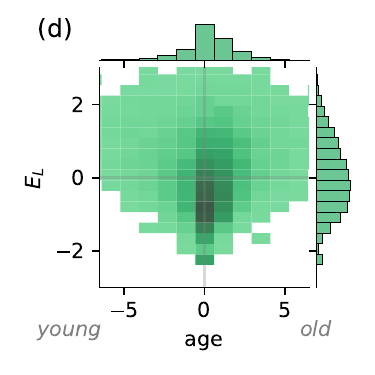}}

\caption{
Joint distributions of the 4 sociodemographic scores (x-axes) and the z-scores of the Reddit engagement (y-axis) of our population.
The sociodemographic features of the users are proxies computed as the dot product between the vector of subreddit users participate in long-term and the sociodemographic scores of the subreddits.
These values are different from $D$, which is a latent variable and is estimated from the model.
A positive correlation is noticeable in (a) and (c), highlighting that users who participate more in more `rich and `female' subreddits tend to comment more.
(a), (b) and (c) are also asymmetric, thus emphasizing that our dataset is biased toward `rich', `left', and `female' users.
}
\label{fig:sociodemo}
\end{figure}

\end{document}